\documentstyle[12pt]{article}

\font\fourteenbf=cmbx12 scaled\magstep1

\newcounter{saveeqn}

\newcommand{\alphaeqn}{\refstepcounter{equation}\setcounter{saveeqn}{\value{equation}}%
\setcounter{equation}{0}%
\renewcommand{\theequation}{%
        \mbox{\arabic{saveeqn}.\alph{equation}}}}%

\newcommand{\reseteqn}{\setcounter{equation}{\value{saveeqn}}%
\renewcommand{\theequation}{\arabic{equation}}}

\begin{document}

\newcommand{\lsi}{\raise0.3ex\hbox{$<$\kern-0.75em\raise-1.1ex\hbox{$\sim$}}}
\newcommand{\gsi}{\raise0.3ex\hbox{$>$\kern-0.75em\raise-1.1ex\hbox{$\sim$}}}
\newcommand{\lsim}{\mathop{\lsi}}
\newcommand{\gsim}{\mathop{\gsi}}
\newcommand{\deltav}{\delta^{(S^2)}}
\newcommand{\gvec}[1]{\mbox{\boldmath $#1$}}
\renewcommand{\vec}[1]{{\bf #1}}
\newcommand{\ptr}{{P_{\rm t}}}
\newcommand{\half}{\mbox{$\frac12$}}
\newcommand{\im}{{\rm Im}}
\newcommand{\re}{{\rm Re}}

\newcommand{\mmdebye}{m^2_{\rm D}}
\newcommand{\mdebye}{m_{\rm D}}
\newcommand{\tr}{_{\rm t}}
\newcommand{\trace}{{\rm tr}}
\newcommand{\disc}{{\rm disc}}

\newcommand{\mref}[1]{(\ref{#1})}
\newcommand{\mlabel}[1]{\label{#1}}

\newcommand{\eq}[1]{Eq.~\mref{#1}}

\newcommand{\hep}[1]{ [#1]}

\begin{titlepage}
\begin{flushright}
\end{flushright}
\begin{centering}
\vfill

{\fourteenbf \centerline{ 
A local Langevin equation for slow long-distance modes 
}
\vskip 2mm \centerline{ 
of hot non-Abelian gauge fields
 }}

\vspace{1cm}

Dietrich B\"odeker \footnote{e-mail: bodeker@bnl.gov}

\vspace{.6cm} { \em Physics Department, 
Brookhaven National Laboratory and  \\ RIKEN BNL Research Center,
Upton, NY 11973 
}

\vspace{2cm}
 
{\bf Abstract}

\vspace{0.5cm}

\end{centering}

\noindent
The effective theory for the dynamics of hot non-Abelian gauge fields
with spatial momenta of order of the magnetic screening scale $ g ^ 2
T $ is described by a Boltzmann equation.  The dynamical content of
this theory is explored. There are three relevant frequency scales, $
g T $ , $ g ^ 2 T $ and $ g ^ 4 T $, associated with plasmon
oscillations, multipole fluctuations of the charged particle
distribution, and with the non-perturbative gauge field dynamics,
respectively.  The frequency scale $ g T $ is integrated out. The
result is a local Langevin-type equation. It is valid to leading order
in $ g $ and to all orders in $ \log(1/g) $, and it does not suffer
from the hard thermal loop divergences of classical thermal Yang-Mills
theory.  We then derive the corresponding Fokker-Planck equation, which  
is shown to generate an equilibrium distribution 
corresponding to 3-dimensional Yang-Mills theory plus a Gaussian 
free field.

\vspace{0.5cm}\noindent


\vspace{0.3cm}\noindent
 
\vfill \vfill
\noindent
 
\end{titlepage}

At sufficiently high temperature $ T $ the running gauge coupling $ g
= g(T) $ in non-abelian gauge theories is small. Nevertheless, long
distance modes of hot non-abelian gauge fields are strongly coupled.
This leads to the phenomenon of magnetic screening on the length scale
$ (g ^ 2 T)^{-1} $. \footnote{The units are chosen such that $ \hbar
= c = k _ {\rm B} =1 $. Furthermore, it is assumed that fermion masses
and chemical potentials can be neglected relative to $ T $.}  It
distinguishes non-abelian from abelian plasmas in which magnetic
fields can be correlated over arbitrarily large distances.

The strong coupling is caused by the large amplitudes of the infrared
gauge field modes. It can be easily understood as follows.  The long
wavelength ($ \lambda \gg T ^{-1} $) modes obey the classical
Rayleigh-Jeans law which states that the energy density is
proportional to $ T $. Then, by dimensional analysis, the magnetic
energy density $ \vec B ^ 2 _ \lambda $ due to wavelengths of order $
\lambda $ must be of order $T \lambda ^{-3} $. This corresponds to $
\vec B _ \lambda \sim T ^{1/2} \lambda ^{-3/2} $ and to a vector
potential $ \vec A _ \lambda \sim T ^{1/2} \lambda ^{-1/2} $. The
Yang-Mills equations of motion are non-linear since they contain
covariant derivatives. When $ \lambda $ approaches $ (g ^ 2 T) ^{-1}
$, both terms in the spatial covariant derivative $ \vec D = \nabla -
g \vec A $ become of order $ g ^ 2 T $, and the term $ g \vec A $ can
no longer be treated as a perturbation.

This breakdown of perturbation theory was first noted in the
thermodynamics of hot non-abelian gauge fields \cite{linde}. The
non-perturbative physics can be ascribed to euclidean pure Yang-Mills
theory in 3 dimensions which is obtained from the 4-dimensional
thermal field theory in the imaginary time formalism by integrating
out all modes with characteristic wavelengths shorter than $(g ^ 2 T)
^{-1}$ \cite{dr}. The 3-dimensional theory can be easily treated
non-perturbatively on a lattice.

Dynamical quantities, which are determined by the real (Minkowski)
time evolution of the $ \lambda \sim (g ^ 2 T) ^{-1} $ modes, are more
difficult to deal with.  An important example  is
the so called hot sphaleron rate. In the standard electroweak theory
it determines the rate for anomalous baryon number violation and is
therefore a crucial ingredient in particle physics models which try to
explain the baryon asymmetry of the universe \cite{Rubakov}.

Fortunately, even for dynamical quantities one can use perturbation
theory to integrate out short distance modes to obtain an effective
theory for the non-perturbative long-distance dynamics.  The first
step is to integrate out ``hard'' physics associated with virtual
momenta of order $ T $. At leading order in $ g $ one obtains the so
called hard thermal loop effective theory \cite{htl,Blaizot:93}. It
describes gauge fields with $ \lambda \gsim (g T)^{-1} $ interacting
with classical colored particles. These particles correspond to quanta
of the $ |\vec k| \sim T $ field modes which have virtual momenta of
order $ g T $ or less.  In a second step one can integrate out the
degrees of freedom associated with $ k _ 0, |\vec k| \sim g T $.  The
resulting effective theory is described by the classical field equations
of motion \cite{langevin}
\alphaeqn%
\mlabel{maxwellboltzmann}
\begin{eqnarray}
        \mlabel{maxwell} 
        D_\mu F^{\mu\nu} &=& \mmdebye W^\nu 
        ,
 \\  
         ( C + v \cdot D)  W &=& \vec{v}\cdot\vec{E} + \xi
        \mlabel{boltzmann} 
        ,
\end{eqnarray}  
\reseteqn
where $ F^{\mu\nu } _ a = \partial ^ \mu A ^{\nu } _ a - \partial ^
\nu A ^{\mu } _ a + g f _ {abc} A ^{\mu } _ b A ^{\nu } _ c $ is the
non-abelian field strength tensor, and $ v ^ \mu \equiv (1, \vec v) $.
The field $ W(x,\vec v) $ represents the fluctuations of adjoint color
charge due to hard particles with 3-velocity $ \vec v $, $\vec v ^ 2
=1 $ \footnote{For notational simplicity we write all expressions in 3
  spatial dimensions. One has to keep in mind that the effective
  theories discussed here, except Eq.~\mref{langevin}, require
  regularization, for example by a continuation to $ d = 3 -2 \epsilon
  $ dimensions.}.  For other approaches leading to the Boltzmann
equation (\ref{boltzmann}) which do not make use of the hard thermal
loop effective theory, see \cite{asy2}-\cite{Blaizot:boltzmann}.  
The current on
the rhs of Eq.~(\ref{maxwell}) is given by
\begin{eqnarray}
        W^\nu (x)\equiv
        \int_{\vec v} v^\nu W(x, \vec  v) 
        \mlabel{wnu}
        ,
\end{eqnarray}  
where $ \int_{\vec v} \equiv \int {d\Omega_\vec{v}}/(4 \pi ) $, times
the square of the leading order Debye mass $ m_{\rm D} \propto g
T$. The lhs of Eq.~(\ref{boltzmann}) contains the linear collision
term
\begin{eqnarray}
        C W (x, \vec v) \equiv \int_{\vec{v'}}
        C(\vec v, \vec v') W (x, \vec v') 
        \mlabel{coll} 
        ,
\end{eqnarray}
which is due to scattering of hard particles with velocities $ \vec v
$ and $ \vec v ' $ with a momentum transfer of order $ g T $.  The
fields in Eq.~\mref{maxwellboltzmann} describe fluctuations close to
thermal equilibrium on length scales larger than $ (gT)^{-1} $. The
collision term, breaking time reflection invariance, is
dissipative. It is accompanied by the Gaussian white noise $ \xi $,
which has vanishing expectation value. Its only non-trivial
correlation function is
\begin{eqnarray}
        \langle \xi _ a(x,\vec v) \xi_b(x',\vec v ') \rangle =
        \frac{2T}{\mmdebye} C(\vec v, \vec v ') \delta_{ab} \delta^4(x
        - x') \mlabel{xi.correlator}
        .
\end{eqnarray}

The collision kernel $ C(\vec v, \vec v ') $ is of order $ g ^ 2 T $.
Its precise form is not important for the purpose of this letter. We
will only use the fact that $ C $ commutes with rotations in $ \vec v
$-space. Thus if $ W (x, \vec v) $ is expanded in spherical harmonics,
the collision operator becomes diagonal and its eigenvalues $ c _ l $ only
depend on $ l $, i.e.,
\begin{eqnarray}
        \int _ {\vec v '} C(\vec v, \vec v') 
        \sum _{lm} W_{lm}(x) Y_{lm}(\vec v')
        = 
        \sum _{lm} c _ l W_{lm}(x) Y_{lm}(\vec v)
        \mlabel{diagonal}
        .
\end{eqnarray} 
$ C $ consists of a piece which depends logarithmically on the
infrared cutoff for the $ |\vec k| \sim g T $ modes 
and of a cutoff independent part.  The $ l = 0 $ eigenvalue $ c _ 0 $
vanishes. For the logarithmic part this was shown in
Ref.~\cite{langevin}. In Ref.~\cite{ladder} it was found that this
holds for the complete $ c _ 0 $ if one uses dimensional
regularization to define $ C $. The vanishing of $ c _ 0 $ ensures
that Eq.~(\ref{boltzmann}) is consistent with Eq.~(\ref{maxwell}),
i.e., that the current on the rhs of Eq.~\mref{maxwell} is conserved.
The complete $ l =1 $ eigenvalue $ c _ 1 $ in dimensional
regularization was explicitly calculated in Ref.~\cite{ay.nll}.

At leading order in $ \log(1/g) ^{-1} $ Eq.~\mref{maxwellboltzmann}
can be approximated by the Langevin equation \cite{langevin}
\begin{eqnarray}
  \vec D \times \vec B = \gamma \vec E + \gvec{\zeta}
  \mlabel{langevin}
        ,
\end{eqnarray} 
where $ E ^ i = F ^{i0} $ and $ B ^ i = -\frac12 \epsilon ^{ijk} F
^{jk} $ are the non-abelian electric and magnetic fields.  The
damping coefficient or color conductivity $ \gamma $ is proportional to  $
T/\log(1/g)$.  The Gaussian white noise $ \gvec{\zeta} $ satisfies
\begin{eqnarray} 
        \langle \zeta ^ i _ a ( x) \zeta ^ j _ b ( x') \rangle
        = 2 \gamma T \delta _{ab}  \delta ^ {ij} \delta (x-x')
        .
\end{eqnarray} 
Therefore it keeps the gauge fields in thermal equilibrium at
temperature $ T $.  Eq.~(\ref{langevin}) implies that the
characteristic frequency of the gauge fields is of order $ \log(1/g) g
^ 4 T $ \cite{frequency}.  

One motivation for integrating out the momentum scale $ |\vec k| \sim
g T $ was that the Hard Thermal Loop effective theory contains UV
divergences which cannot be removed by renormalization.  Thus
simulations of the non-perturbative gauge field dynamics using the HTL
effective theory do not have a continuum limit.  This problem still
persists in (\ref{maxwellboltzmann}), however. If one would try to
take the continuum limit of (\ref{maxwellboltzmann}) by sending the UV
cutoff to infinity one would encounter precisely the same divergences,
simply because the collision term and the noise, which distinguish
Eq.~(\ref{maxwellboltzmann}) from the HTL effective theory, can be
neglected at very large momenta as they do not grow as fast as the
derivative terms.

The effective theory (\ref{langevin}), on the other hand, is UV finite
\cite{asy2}. Thus it is well suited for non-perturbative lattice
simulations.  It was used to compute the hot sphaleron rate by Moore
\cite{moore.log}. In Ref.~\cite{moore.higgs} it was extended to
account for a Higgs field with a thermal mass of order $ g ^ 2 T $.

Recently Arnold obtained a non-local Langevin equation for the $ k _ 0
\sim g ^ 4 T $ dynamics which is valid to leading order in $ g $ and
to all orders in $ \log(1/g)^{-1} $ \cite{arnold.langevin}. Arnold and
Yaffe showed that it can be used to systematically improve the theory
\mref{langevin} in a perturbative expansion in $ \log(1/g)^{-1} $
\cite{ay.nll}. They found that Eq.~\mref{langevin} is still valid at
next-to-leading order in $ \log(1/g) ^{-1} $ if one includes a
next-to-leading log correction in the color conductivity $ \gamma $
\cite{ay.nll}.  With this correction Moore's  result for the hot sphaleron
rate \cite{moore.log} agrees surprisingly well with different
simulations of the hard thermal loop effective theory
\cite{particles}-\cite{bmr}, which include all orders in $ \log(1/g)^{-1}
$ but which do not have a continuum limit \cite{bms}.

The purpose of this letter is to fully explore the dynamical content
of Eq.~(\ref{maxwellboltzmann}). It will be shown that, in addition to
the well known plasmon oscillations and the non-perturbative gauge
field dynamics, Eq.~(\ref{maxwellboltzmann}) describes fluctuations of
multipole moments of $ W $ with a characteristic frequency of order $
g ^ 2 T $. Then we will obtain a generalization of
Eq.~(\ref{langevin}) by integrating out the physics of plasmon
oscillation which is characterized by the frequency scale $ g T $. The
result, Eq.~(\ref{maxwellboltzmann.2}), is a Langevin equation
which is local in space and time. In contrast to Eq.~(\ref{langevin})
and to the Langevin equation of Ref.~\cite{arnold.langevin} it contains
two different frequency scales, $ g ^ 2 T $ and $ g ^ 4 T $.

In the following power counting estimates logarithms of $ g $ will be
ignored.  All approximations will be valid at leading order in $ g $
and all orders in $ \log(1/g)^{-1} $. Sometimes it will be convenient
to write the scalar, vector, and multipole components of $ W $
separately. The latter will be denoted by $ \widetilde{W} $, so that
\begin{eqnarray}
        W(\vec v ) = W ^ 0 + 3 \vec v \cdot \vec W + \widetilde{W} (\vec v)
        ,
\end{eqnarray} 
where the factor 3 simply follows from the definition \mref{wnu} and from
$ \int _ {\vec v} v ^  i v ^  j = \frac13 \delta ^ {ij} $.

The only length scale in Eq.~(\ref{maxwellboltzmann}) is set by $ C
^{-1} $ and the magnetic screening length which are both of order $ (g
^ 2 T) ^{-1}$.  We have already seen that the covariant derivative $
\vec D $ is then of the same order of magnitude as the ordinary
derivative.  Therefore one can obtain important information about
the dynamical content of Eq.~(\ref{maxwellboltzmann}) already
by considering the linearized equations of motion. We Fourier transform them, 
\alphaeqn
\mlabel{linear}
\begin{eqnarray}
        i k^0 \vec E + i \vec k \times \vec B
        &= & \mmdebye \vec W 
        \mlabel{ampere.linear}
        ,
\\
        i\vec k \cdot \vec E &=& \mmdebye W^0 
        \mlabel{gauss.linear}
        ,
\\ 
        (C -i v\cdot k) W &=& \vec v \cdot \vec E
        + \xi   
        \mlabel{boltzmann.linear}
        ,
\end{eqnarray}
\reseteqn 
and consider only $ |\vec k | \sim g ^ 2 T $.  The magnetic field can
be eliminated from Eq.~\mref{ampere.linear} using $ \vec k \times \vec
E = k^0 \vec B $,
\begin{eqnarray}
        i k_0 \vec E 
       + \frac{i}{k^0} \vec k\times   
        \vec k \times \vec E 
        &= & \mmdebye \vec W 
        \mlabel{ampere.gT}
        .
\end{eqnarray}

First consider the case $ k _ 0 \gg |\vec k| $.
Then one can neglect the second term on the lhs of
Eq.~(\ref{ampere.gT}), so that
\begin{eqnarray}
        \vec E = -i \frac{\mmdebye }{k^0}  \vec W 
        \mlabel{E}
        .
\end{eqnarray}
Combining Eqs.~(\ref{gauss.linear}) and (\ref{E}) one finds that $ W ^
0 \sim \vec k \cdot \vec W/k ^ 0 \ll |\vec W| $. Thus one can neglect
$ W ^ 0 $ in Eq.~\mref{boltzmann.linear}.  Since $ k _ 0 \gg C \sim g
^ 2 T $ one can neglect the collision term and the noise in
Eq.~\mref{boltzmann.linear} and one can approximate $ v \cdot k \simeq
k _ 0 $, which gives
\begin{eqnarray}
         k _ 0 \left( 3 \vec v \cdot \vec W + \widetilde{W} \right)  
        =   \frac{m^2_{\rm D} }{k^0}  \vec v \cdot \vec W
      \mlabel{boltzmann.gT}
      . 
\end{eqnarray}
If one multiplies Eq.~(\ref{boltzmann.gT}) with $ \vec v $ and integrates
over $ \vec v $, $ \widetilde{W} $ drops out and one obtains
\begin{eqnarray}
        \left( k _ 0 ^ 2 - \frac13 m^2_{\rm D}  \right) \vec W=0
        \mlabel{plasmon}
        .
\end{eqnarray}
Thus $ \vec W $ oscillates with the plasmon frequency $
\omega_{\rm pl} = \frac {1}{\sqrt{3}} m_{\rm D}$.  The plasmon
oscillations also involve the electric field which is determined by
Eq.~\mref{E}, and $ W ^ 0 $ which is obtained from $ \vec E $ using
Eq.~(\ref{gauss.linear}).  For $ \widetilde{W} $, \eq{boltzmann.gT}
would imply $ \widetilde{W} \propto \delta(k ^ 0)$, meaning that $
\widetilde{W} $ is time independent. But Eq.~(\ref{boltzmann.gT}) is only
an approximation valid when $k^0 \gg g ^ 2 T$. Thus from
Eq.~(\ref{boltzmann.gT}) one can only conclude that $ \widetilde{W} $
evolves more slowly than $W^\mu$.  We will now see that the
characteristic frequency of $\widetilde{W} $ is of order $g^2 T$.

The appearance of the frequency scale $ g ^ 2 T $, which so far has
not been discussed in the literature, is immediately obvious if we
choose the $ z $-axis in Eq.~(\ref{linear}) parallel to $ \vec k $ and
expand $ W(\vec v) $ in spherical harmonics $ Y_{lm}(\vec v)$. Then
the factor $ v\cdot k $ in \mref{boltzmann.linear} equals $ k ^ 0 -
|\vec k| \cos \theta _ {\vec v} $ and does not mix $ W _{lm} $ with
different $ m $.  Furthermore, the collision term is diagonal (cf.\
Eq.~(\ref{diagonal})), so that the $ W _{lm} $ with $ |m|\ge 2 $ are
completely decoupled from the gauge fields. The dynamics of these
modes is thus governed by Eq.~\mref{boltzmann.linear} but without the
electric field. They perform oscillations with frequencies determined
by $ \vec v \cdot \vec k \sim g ^ 2 T $, which are driven by the noise
term $ \xi $, and which are damped by the collision term at a rate of
order $ C \sim g ^ 2 T $.

To include the $ |m|\le 1 $ modes of $ W $ and the gauge fields in
this picture it is convenient to solve Eq.~(\ref{boltzmann.linear})
for $ W $,
\begin{eqnarray}
        W(\vec v) = \int _{\vec v'}  
         G _ k (\vec v, \vec v')  
        \Big[   \vec v' \cdot \vec E + \xi (\vec v')
        \Big]
        \mlabel{W}
        .
\end{eqnarray}  
Here $ G _ k (\vec v, \vec v') $ denotes the $ \vec v $-space inverse
of the operator $ C - i v \cdot k $. It has cuts in the lower half of
the complex $ k ^ 0 $-plane which reflects the damping caused by the
collision term. To understand the role of the electric field in
Eq.~(\ref{W}) we insert this result into Eq.~\mref{ampere.gT},
\begin{eqnarray}
        i k_0 \vec E 
        + \frac{i}{k^0} \vec k \times \vec k \times \vec
         E 
        =  
        \mmdebye \int _{\vec v,\vec v '} 
        \vec v G _ k (\vec v, \vec v') 
        \Big[ \vec v'
         \cdot \vec E + \xi (\vec v') \Big] 
        \mlabel{ampere.g2T} 
        .
\end{eqnarray} 
Since we consider $ k _ 0 \sim |\vec k| \sim g ^ 2 T $, both terms on
the lhs of Eq.~(\ref{ampere.g2T}) are of order $ g^{2} T^{} \vec E $, while
$ \mmdebye G _ k \vec E $ on the rhs is of order $ T^{}\vec E $.
Consequently the lhs of Eq.~(\ref{ampere.g2T}) can be neglected
completely when $ k _ 0 \sim g ^ 2 T $. This is a crucial point which
will later allow us to drop the term $ D _ 0 \vec E $ for all
frequencies smaller than $ g T $, even though both terms on the lhs of
Eq.~(\ref{ampere.g2T}) are of the same order of magnitude when $ k ^ 0
\sim g ^ 2 T $. It implies that the two terms on the rhs of
Eq.~(\ref{ampere.g2T}) must cancel.  The electric field is thus
entirely determined by the noise $ \xi $ in such a way that the
current $\mmdebye \vec W$ vanishes at leading order in $ g $.  Gauss'
law (\ref{gauss.linear}) , together with Eq.~\mref{W} gives $ W ^ 0
\sim g ^ 2 \widetilde{W} $. Therefore both $ W ^ 0 $ and $ \vec W $
are small compared to $ \widetilde{W} $ when $ k _ 0 \sim g ^ 2 T $.

Finally we briefly recall the case $ k _ 0 \ll g ^ 2 T $ which has
been studied in great detail
\cite{langevin}, \cite{asy}-\cite{Arnold:subtleties}, 
\cite{arnold.langevin}.  Again,
Gauss' law implies that one can neglect $ W ^ 0 $ in
Eq.~\mref{boltzmann.linear}. Furthermore, one can approximate $ v \cdot
k \simeq -\vec v \cdot \vec k $.  This corresponds to dropping $ D _ 0
W $ in Eq.~\mref{boltzmann}. Then $ W $ is not dynamical, but it is
fixed by the gauge fields and the noise at the same instant of
time. For $ k _ 0 \ll g ^ 2 T $ the ``magnetic'' term $ k _ 0 ^{-1}
\vec k \times \vec k \times \vec E $ in Eq.~(\ref{ampere.g2T}), which
so far has not played any role, becomes relevant.  It is now much larger
than the kinetic term $ k _ 0 \vec E $ which can be neglected. Then
Eq.~(\ref{ampere.g2T}) gives $ k _ 0 \sim g ^ 4 T $ as the
characteristic frequency of the magnetic sector.

To summarize the discussion of the linearized equations of motion
\mref{linear}, we have found that the characteristic frequency of the
electric field $ \vec E $ and the 4-current $ W ^ \mu $ is given by $
\omega _{\rm pl} \sim g T $. The characteristic frequency of $
\widetilde{W} $, i.e., of the $ l \ge 2 $ components of $ W $ is $ k ^
0 \sim g ^ 2 T $. Finally, the characteristic frequency of the
magnetic fields is of order $ g ^ 4 T$.

Now consider the effect of interactions in
Eq.~(\ref{maxwellboltzmann}). Since $ \vec D\sim\nabla $ it is clear
that none of the above order of magnitude estimates is changed by
replacing $ \nabla \to \vec D $. The basic picture of plasmon
oscillations and multipole oscillations is unaffected, except that
they occur in a quasi-static gauge field background.

The effect of replacing $ \partial _ 0 $ by $ D _ 0 $ is less obvious
and it depends on which frequencies are involved.  First consider the
plasmon oscillations. One can estimate the size of $ A ^ 0 $ using $
\nabla A ^ 0 \sim \vec E $. Because of equipartition $ \vec E(x) $ and
$ \vec B (x) $ are of the same order of magnitude. This gives the same
estimate for $ A ^ 0 (x) $ as for $ \vec A (x) $, i.e., $ A ^ 0(x)
\sim g T $.  Consequently one can neglect $ g A ^ 0 $ in the covariant
time derivative acting on  fields with frequencies of order $ g T $, and
Eq.~\mref{plasmon} and the corresponding result for $ \vec E $ are
also valid in the interacting theory. Since Gauss' law contains a
spatial derivative, the result for $ W ^ 0 $ will look different in
the presence of interactions but again the order of magnitude of $ W ^
0 $ is unchanged.

Now consider $k^0 \lsim g^2T$, ignoring for a moment the physics of $
k _ 0 \sim g T $. To estimate $ A ^ 0 $ we need to know the size of $
\vec E (k)$. We have seen that for $k^0 \lsim g^2T$ the electric field
is of the same order of magnitude as $ \xi $. The latter can be
estimated from the Fourier transform of Eq.~\mref{xi.correlator},
which gives $ \xi(k) \sim k _ 0 ^{-1/2} |\vec k| ^{- 3/2} $. Then one
can use $ \nabla A ^ 0 (x) \sim \int d ^ 4 k e ^ {-ik \cdot x }\vec E
(k) $, which gives $ A ^ 0 (x) \sim g ^ 2 T $, $ g ^ 3 T $ for $k^0
\sim g^2T$, $ g^4 T $, respectively.  Note that these estimates are
smaller than the one obtained from the equipartition argument
above. This is because the electric field at a given time is dominated
by Fourier components with frequencies of order $ g T $.
Nevertheless, $ A _ 0 $ is big enough to be able to change the
estimate $ \partial _ 0 \sim g ^ 4 T $ to $ D _ 0 \sim g ^ 3 T $ when
$ D _ 0 $ acts on a field with frequency of order $ g ^ 4 T $. Still,
this would  be a factor $ g $ smaller than $ \vec D \sim g ^ 2 T
$, and neglecting $ D _ 0 $ would still be justified.

We are now in the position to tentatively write down an effective
theory for the physics associated with $ k _ 0 \lsim g ^ 2 T $ which
corresponds to integrating out $ \vec E $ and $ W ^ \mu $ as
dynamical degrees of freedom. We have seen that for both $k^0 \sim
g^2T$ and $k^0 \sim g^4T$ one can neglect the term $D_0 \vec E$. The
spatial components of \eq{maxwell} can thus be replaced by \alphaeqn
\mlabel{maxwellboltzmann.2}
\begin{eqnarray}
        \vec D \times \vec B = \mmdebye \vec W 
        \mlabel{maxwell.2}
        .
\end{eqnarray}
In Eq.~\mref{boltzmann} we were able to neglect $W^0$ both for $k^0
\sim g^2T$ and $k^0 \sim g^4T$. Therefore one can drop $ W ^ 0 $
altogether. We have also seen that for $k^0 \sim g^2T$ we could
neglect $\vec W$. For $k^0 \sim g^4T$ we were able to neglect all time
derivatives on the lhs of \mref{boltzmann}, and in particular the term
$D_0 \vec W$. Thus for both $k^0 \sim g^2T$ and $k^0 \sim g^4T$ the
term $D_0 \vec W$ can be neglected and we can replace \eq{boltzmann}
by
\begin{eqnarray}
        3(c_1 + \vec v \cdot \vec D)
        \vec v \cdot \vec W
        + (C + v \cdot D) \widetilde{W}  = \vec v \cdot \vec E + \xi
        \mlabel{boltzmann.2}
        ,
\end{eqnarray}
\reseteqn
where $c_1$ is the $l=1$ eigenvalue of the collision operator $ C
$. Eq.~(\ref{maxwell.2}) is no longer a dynamical equation. Instead,
it fixes the 3-current in terms of the gauge fields at the same
instant of time.

The only remaining question is whether plasmon oscillations affect the
low frequency ($ k ^ 0 \lsim g ^ 2 T $) dynamics through interactions.
To address this issue we consider the gauge field polarization tensor
$ \Pi ^{\mu \nu }(k) $ in the theory \mref{maxwellboltzmann} at one
loop. We are interested in  $ k _ 0 \lsim g ^
2 T $,
and loop momenta with $ q ^ 0 \sim g T $.
Inside the loop one can neglect the effects of $ C $ and $ \xi $
since $ q ^ 0 \gg g ^ 2 T $. Without these terms
Eq.~(\ref{maxwellboltzmann}) has the same form as the non-abelian
Vlasov equations \cite{Blaizot:93} which describe the hard thermal
loop effective theory. Therefore the calculation of $ \Pi ^{\mu \nu
}(k) $ is precisely the same as the one in Ref.~\cite{ladder} where it was
found that the leading order in $ g $ contribution is due to
space-like loop momenta. This shows that the time-like loop momenta $
q ^ 0 \sim g T $, $ |\vec q| \sim g ^ 2 T $ considered here do not
contribute at leading order in $ g $.

Written in $ A ^ 0 = 0 $ gauge \footnote{Note that we have
not assumed this gauge when deriving Eq.~(\ref{maxwellboltzmann.2}).},
Eq.~(\ref{maxwellboltzmann.2}) is a Langevin equation which is purely
dissipative, i.e., it contains only first order time derivatives of
the dynamical degrees of freedom $ \vec A $ and $ \widetilde{W} $,
\begin{eqnarray}
        \frac{ \partial }{\partial t} 
        \left( \vec v \cdot \vec A + \widetilde{W}  \right) 
        = 
        - (C + \vec v \cdot \vec D) 
        \left( 3 \vec v \cdot \vec W + \widetilde{W}  \right)  
        + \xi
        \mlabel{langevin.tag}
        .
\end{eqnarray}

For the subsequent discussion I remind the reader of some features of
a Langevin equation for some set of degrees of freedom $ \varphi _ \alpha $
\cite{zinn-justin},
\begin{eqnarray}
                \frac{ \partial }{\partial t} \varphi _ \alpha (t) 
        =
        - f _ \alpha [\varphi(t)] + \xi _ \alpha  (t)
        \mlabel{langevin.generic}
        ,
\end{eqnarray} 
where $ f $ is a functional of $ \varphi $ at time $ t $.  In the
present context the index $ \alpha $ would represent the spatial
coordinates, $ \vec v $ and vector as well as color indices.  Both
Eq.~(\ref{langevin}) in $ A ^ 0 = 0 $ gauge, and
Eq.~(\ref{langevin.tag}) are of this form, as well as the Langevin
equation of Ref.~\cite{arnold.langevin}.  In our case $ f $ is a {\em
local} functional of $ \varphi $, that is, it contains only the $
\varphi _ \alpha $ and a finite number of spatial derivatives. In
contrast, in the Langevin equation of Ref.~\cite{arnold.langevin} $ f
$ is a spatially non-local functional of the gauge fields \footnote{In
order to avoid confusion I would like to stress the following. By
reintroducing the $ W $-field the Langevin equation of
Ref.~\cite{arnold.langevin} can be written in a local form. Then,
however, the equation for $ W $ would be a {\em constraint} and not an
equation of motion.} . Solving Eq.~(\ref{langevin.generic}) for any
given realization of the noise $ \xi $ one generates an ensemble of
field configurations for any time $ t $. If the noise is Gaussian and
white (that is, it has a frequency independent spectrum),
\begin{eqnarray}        
        \langle \xi _ \alpha (t) \xi _ \beta (t') \rangle = 
        2 T \Omega _{\alpha \beta} \delta (t-t')
	\mlabel{white}
	,
\end{eqnarray} 
the probability distribution for field configurations $ P(\varphi, t)
$ satisfies the Fokker-Planck equation
\begin{eqnarray} 
        \frac{ \partial }{ \partial \varphi _ \alpha}
        \left[ f _ \alpha + T \Omega _{\alpha \beta}
        \frac{ \partial }{ \partial \varphi _ \beta} \right] 
        P = 0
        \mlabel{fp}
        .
\end{eqnarray} 
An important case is that $ f $ is related to the derivative of a
Hamiltonian $ H $ through
\begin{eqnarray}
	f _ \alpha =
	\Omega _{\alpha \beta} \frac{ \partial H_{} }{ \partial \varphi _
	\beta }
	\mlabel{f}
	.
\end{eqnarray} 
Then the thermal equilibrium distribution $ P_{\rm eq} = \exp[-H_{\rm
}/T] $ is a stationary solution of the Fokker-Planck
equation. Furthermore, for any initial configuration $ P(\varphi, t) $
approaches $ P_{\rm eq} $ for large times.

For Eq.~(\ref{langevin}) we have $ \Omega _{\alpha \beta} = \gamma ^{-1} 
\delta _{\alpha \beta} $ and $ f _ \alpha = \Omega _{\alpha \beta}
(\partial H_{\rm 3d} / \partial \varphi _ \beta) $ with the
Hamiltonian
\begin{eqnarray}
        H_{\rm 3d} = 
        \frac12 \int d ^ 3 x  \vec B ^ 2 
        \mlabel{h3d}
        .
\end{eqnarray}
Thus for Eq.~(\ref{langevin}) the probability distribution approaches
$ \exp[-H_{\rm 3d}/T] $ for large times. Note that 
$ H_{\rm 3d}/T $ is the action of magnetostatic Yang-Mills theory
\cite{braaten.nieto} which is obtained for equilibrium quantities by
dimensional reduction and by integrating out the $ A _ 0 $ field.

Beyond the leading and next-to-leading log approximation, for which
Eq.~(\ref{maxwellboltzmann}) reduces to Eq.~(\ref{langevin}), it has
so far not been understood whether Eq.~(\ref{maxwellboltzmann})
reproduces the correct thermodynamics of long distance Yang-Mills
fields.   The
noise correlator of Ref.~\cite{arnold.langevin} depends   
on the gauge fields, and is therefore not of
the form (\ref{white}). With a field dependent noise correlator the
time discretisation of the Langevin equation is ambiguous. In
Ref.~\cite{arnold.langevin} this ambiguity was fixed by {\em
postulating} that the Langevin equation yields the equilibrium
distribution $ \exp[- H_{\rm 3d}/T] $.  The considerations of
Ref.~\cite{litim.fluctuations} did not take into account the
non-linear character of Eq.~(\ref{boltzmann}). 

We will now determine the equilibrium distribution which is generated
by Eq.~(\ref{langevin.tag}).  Obviously the rhs of
Eq.~(\ref{langevin.tag}) can not be written like in
Eq.~(\ref{langevin.generic}) with an $ f $ of the form (\ref{f}).  It
is well known, however, that the equilibrium distribution does not
specify the function $ f _ \alpha $ uniquely \cite{zinn-justin}. One
may add an extra term $ F _ \alpha $,
\begin{eqnarray} 
         f _ \alpha = \Omega _{\alpha \beta} \frac{ \partial H }{ \partial
                \varphi _ \beta} + F _ \alpha
        \mlabel{f.generic}
	,	
\end{eqnarray}
without changing the equilibrium distribution provided that $ F _
\alpha $ satisfies
\begin{eqnarray}
        \frac{ \partial F _ \alpha }{\partial \varphi _ \alpha}
        = F _ \alpha  \frac{ \partial H }{ \partial
                \varphi _ \alpha }
        \mlabel{condition}
        .
\end{eqnarray} 
We will see that our Langevin equation (\ref{langevin.tag}) is
indeed of the form (\ref{langevin.generic}), (\ref{f.generic}) with
an $ F $ satisfying
the condition (\ref{condition}). From Eq.~(\ref{xi.correlator}) we see
that we have to identify
\begin{eqnarray}
        \Omega _{\alpha \beta} \leftrightarrow 
        \frac{ 1}{m^2_{\rm D} } C(\vec v, \vec v')  \delta(\vec x
        -\vec x') \delta ^{ab} 
	\mlabel{identify}
        .
\end{eqnarray}
The next question concerns the relevant Hamiltonian.
Eq.~(\ref{maxwellboltzmann}) was obtained from the Hard Thermal Loop
effective theory for which the Hamiltonian reads \cite{htlhamiltonian} 
\begin{eqnarray}
        H_ {\rm HTL}&=& \frac12 \int d^3 x  \left\{
        \vec{E}^2 +\vec{B}^2
        + \mmdebye\int_{\vec v}
        W^2 \right \}
        \mlabel{htlhamiltonian}
        .
\end{eqnarray}
The effective theory (\ref{maxwellboltzmann.2}) is obtained from
Eq.~(\ref{maxwellboltzmann}) by integrating out $ \vec E $ and $ W ^
\mu $. Thus one can expect that the relevant Hamiltonian is obtained
from (\ref{htlhamiltonian}) by dropping these fields, i.e.,
\begin{eqnarray}
        H_ {\rm }&=& \frac12 \int d^3 x  \left\{
        \vec{B}^2
        + \mmdebye\int_{\vec v}
        \widetilde{W} ^2          \right \}
        \mlabel{hamiltonian}
        .
\end{eqnarray}
Using
\begin{eqnarray}
        \frac{ \delta H }{ \delta
                \vec A} = \vec D \times \vec B, \quad 
        \frac{ \delta H }{ \delta
                \widetilde{W}  } = m^2_{\rm D} \widetilde{W}
	, 
\end{eqnarray}
together with Eq.~(\ref{identify}) we indeed obtain the terms on the
rhs of  Eq.~(\ref{langevin.tag}) which contain the collision term. 
The remaining terms on the rhs of Eq.~(\ref{langevin.tag}) have to be
identified with 
$ F _ \alpha $,
\begin{eqnarray} 
	F _ \alpha \leftrightarrow  - \vec v \cdot \vec D 
	\left( 3 \vec v \cdot \vec W + \widetilde{W}
	\right)
	.
\end{eqnarray}  
We will now see that for this $ F _ \alpha $ both the lhs and the rhs
of Eq.~(\ref{condition}) are zero, so that Eq.~(\ref{condition}) is
indeed satisfied.  First consider the lhs of
Eq.~(\ref{condition}). When the $ \varphi $-derivative acts on the
gauge field contained in $ \vec v \cdot \vec D $ one obtains zero due
to the contraction of color indices and the antisymmetry of the
structure constants. When it acts on $ 3 \vec v \cdot \vec W +
\widetilde{W} $ the result is an integral over a total spatial
derivative which again vanishes. The rhs of Eq.~(\ref{condition}) can
be written as
\begin{eqnarray}
        -\frac{ 1}{m_{\rm D} ^ 4} \int d ^ 3 x \int _ {\vec v}
        \frac{ \delta H }{\delta \varphi (x, \vec v)}
        \vec v \cdot \vec D  \frac{ \delta H }{\delta \varphi (x, \vec v)}
        \nonumber 
        ,
\end{eqnarray}
with $ \varphi  \equiv \vec v \cdot \vec A + \widetilde{W} $.
This again is an integral of a total derivative and vanishes.

We conclude that the probability distribution generated by the
Langevin equation (\ref{langevin.tag}), with $ \vec W $  given
by Eq.~(\ref{maxwell.2}), approaches for large times the Boltzmann
distribution $ P _ {\rm eq} = \exp(-H/T) $ with the Hamiltonian
(\ref{hamiltonian}). This is not the equilibrium distribution corresponding
to dimensional reduction because it contains the additional
field $ \widetilde{W} $. In $ H $, however, $ \widetilde{W}
$ is not coupled to the gauge fields. Equal time correlation
functions of $ \vec A $ computed with the help of $ H $ are thus the same
as in 3-dimensional Yang-Mills theory.

Another interesting observation is the following.
Without a kinetic term $ D _ 0 \vec E $
there are no propagating gauge field waves in the theory
(\ref{maxwellboltzmann.2}). These would cause the same non-local UV
divergences as in classical thermal Yang-Mills theory \cite{bms},
which are still present in Eq.~(\ref{maxwellboltzmann}) since the
effect of the $ W $ fields can be neglected in the
ultraviolet. Therefore the UV divergences of
Eq.~(\ref{maxwellboltzmann.2}) can be expected to be local.

To summarize, we have found that Eq.~(\ref{maxwellboltzmann})
describes plasmon oscillations, multipole fluctuations of color
charge, and the non-perturbative gauge field dynamics. These processes
are associated with characteristic frequencies $ g T $, $ g ^ 2 T $,
and $ g ^ 4 T $, respectively.  An effective theory
(\ref{maxwellboltzmann.2}) was constructed which reproduces the slow
($ k _ 0 \lsim g ^ 2 T $) dynamics at leading order in $ g
$. Previously \cite{langevin,eff,arnold.langevin,ay.nll} it was
implicitly assumed that all modes with $ k _ 0 \gg g ^ 4 T $ decouple
from the non-perturbative gauge field dynamics.  Here it was shown
that this is indeed the case for the plasmon oscillation.  To see
whether the same is true for the multipole fluctuations requires a
more detailed analysis.

{\bf Acknowledgements} I would like to thank P.\ H.\ Damgaard, M.\
Laine, G.D.\ Moore and K.\ Rummukainen for useful discussions and
suggestions. This work was supported by the TMR network ``Finite
temperature phase transitions in particle physics'', EU contract
ERBFMRXCT97-0122.




\begin{thebibliography}{99}
\bibitem{frequency} The lhs of Eq.~(\ref{langevin}) can be estimated
as $ (g ^ 2 T )^{2} \vec A $, while on the rhs $ \gamma \vec E \sim
T/\log(1/g) k _ 0 \vec A $. This gives $ k _ 0 \sim \log(1/g) g ^ 4 T
$.
\end{thebibliography}
\end{document}